\documentclass[english,aps,preprint]{revtex4}
\usepackage[T1]{fontenc}
\usepackage[latin9]{inputenc}
\usepackage{amsmath}
\usepackage{amssymb}
\usepackage{graphicx}

\makeatletter
\@ifundefined{textcolor}{}
{%
 \definecolor{BLACK}{gray}{0}
 \definecolor{WHITE}{gray}{1}
 \definecolor{RED}{rgb}{1,0,0}
 \definecolor{GREEN}{rgb}{0,1,0}
 \definecolor{BLUE}{rgb}{0,0,1}
 \definecolor{CYAN}{cmyk}{1,0,0,0}
 \definecolor{MAGENTA}{cmyk}{0,1,0,0}
 \definecolor{YELLOW}{cmyk}{0,0,1,0}
}

\@ifundefined{showcaptionsetup}{}{%
 \PassOptionsToPackage{caption=false}{subfig}}
\usepackage{subfig}
\makeatother

\usepackage{babel}
\begin{document}

\title{Dynamics of the relativistic Gross-Pitaevskii equation with harmonic
potential: Following the variational approach}

\author{F. J. Poveda-Cuevas}
\email{fjpovedac@gmail.com}

\affiliation{Instituto de F\'{i}sica, Universidad Nacional Aut\'{o}noma de M\'{e}xico,
Apartado Postal 20-364, 01000 México D. F., M\'{e}xico}

\author{R. P. Teles}
\email{rafael.fisica.unesp@gmail.com}

\affiliation{Department of Physics and Astronomy, Rice University, 6100 Main St.,
Houston, Texas 77251, USA}
\begin{abstract}
The role of the collective excitations as well as the free expansion
dynamics provide a key diagnostic tools for trapped Bose-Einstein
condensations. Based on such dynamics we proposed to study the relativistic
version of them in the context of a macroscopic occupation of the
ground-state for spin-$0$ particles. Therefore we used the Higgs
model where the external trap is introduced by a non-minimal coupling.
Along with variational method, we obtained a nonlinear coupling between
dipolar and monopolar modes. Furthermore, the free expansion is no
longer ballistic reaching a relativistic confinement.
\end{abstract}
\maketitle

\section{introduction}

Phenomena such as superfluidity, superconductivity or topological
defects were successfully described by theories of nonlinear fields,
and the correspondence between field theories for non-relativistic
and relativistic systems has appeared throughout history. For example,
the two-fluid model of the Ginzburg-Landau theory \cite{Landau-physrev60,Abrikosov-sovphysJETP5}
has an analogy with the scalar electrodynamics \cite{Nielsen-nuclphysb61,Nambu-prd10,tHooft-nuclphysb190}.
There is also a clear analogy between the nonlinear Schrödinger equation
\cite{Rajaraman-book1982} with Abelian Higgs model (AHM) for spinless
scalar bosons \cite{Peskin-book1995}. An interesting case is the
nonlinearity in the equation of motion that describes the dynamics
of the macroscopic occupation of ground state for many bosons, i.e.
the Gross-Pitaevskii equation (GPE) \cite{Gross-nc20,Pitaevskii-sovphysJETP13,Gross-jmathphys4}.
Thus, the GPE approach preserves some physical properties and characteristics
such as superfluidity and vorticity, which are present in Bose-Einstein
condensation (BEC) of atomic trapped gases \cite{Dalfovo-revmodphys71}.
In fact, during the last decades, the GPE is in agree with the experimental
demonstrations of BEC in harmonically trapped ultracold gases \cite{Anderson-science269,Davis-prl75,Bradley-prl78}.
These ones have\textbf{ }stimulated the interest of many fields in
physics \cite{Bloch-revmodphys80}. The purpose of this paper is to
show a relativistic version of the GPE in a harmonic trap.

There are strong motivations to study the dynamics of GPE in different
scenarios. For instance, the analogy with BEC in the relativistic
regime can be associated to unified cosmological models and observations
related with Dark Energy and Dark Matter \cite{Fukuyama-progthphys115,Das-classquangrav32}.
It is remarkable that the hydrodynamics equations of GPE are recovered
from the Klein-Gordon equation in a simple model for the gravitational
potential \cite{Suarez-prd92}. In addition, there are perspectives
for using experiments of ultracold quantum gases to learn about the
dynamics of high energy particles due to the dynamical universality
classes of many body systems far from equilibrium \cite{Pineiro-prd92}.
Finally, the understanding of quantum criticality is through studying
the finite-density $O\left(2\right)$ model \cite{Hazzard-pra84},
which is important for superfluid-insulator boundary in Bose-Hubbard
model being analog to the equation of motion proposed in this article.

The space-like quadratic term in GPE corresponds to the external harmonic
potential. This is naturally introduced by coupling the internal degrees
of freedom of the atoms with an external field. Actually, GPE can
be considered as a Schrödinger equation for a harmonic oscillator
with an effective interaction term proportional to the density. In
the same spirit, a harmonic potential in the Klein-Gordon equation
(KGE) for a complex scalar field can be introduced by following coupling
(adopting the units $\hbar=c=1$) \cite{Mirza-communthphys42,Chargui-communthphys56}:
\begin{equation}
\hat{p}_{i}\rightarrow\hat{p}_{i}-im\omega_{i}\hat{x}_{i},\,\,\,\hat{p}_{i}^{\dagger}\rightarrow\hat{p}_{i}+im\omega_{i}\hat{x}_{i}.\label{eq:01}
\end{equation}
where $\omega_{i}$ is the external trapping frequency in $i$-direction
($i=1,2,3$). This approach is motivated by the prescription of Moshinsky
for Dirac oscillator \cite{Moshinsky-jpamaththeo22,Benitez-prl64,Dominguez-eurphyslett13,Dominguez-physletta162},
whose coupling preserves the simple formulation and the analytic solution
of the Dirac oscillator \cite{Rao-physscr77,Boumali-physscr84,Boumali-physletta346}.
Now, we simply have a way to coupling an AHM to external harmonic
potential, which is basically a nonlinear KGE for a complex scalar
field, $\hat{\Phi}\left(\mathbf{r},t\right)\equiv\hat{\Phi}$. Thus,
the equation of motion is given by: 
\begin{equation}
\frac{\partial^{2}\hat{\Phi}}{\partial t^{2}}-\nabla^{2}\hat{\Phi}+\sum_{i=1}^{3}\left(m^{2}\omega_{i}^{2}x_{i}^{2}-m\omega_{i}\right)\hat{\Phi}+m^{2}\hat{\Phi}+\lambda\left(\hat{\Phi}^{\dagger}\hat{\Phi}\right)\hat{\Phi}=0,\label{eq:02}
\end{equation}
similarly for $\hat{\Phi}^{\dagger}$. Note that when $\omega_{i}\rightarrow0$
we recover the AHM. This eq.\eqref{eq:02} has essentially the same
form of GPE except by second order derivative in time, therefore we
call it as Relativistic Gross-Pitaevskii equation (RGPE). We approximate
the quantum field to classical fields: $\hat{\Phi}=\Phi+\delta\hat{\Phi}$
and $\hat{\Phi}^{\dagger}=\Phi^{\ast}+\delta\hat{\Phi}^{\dagger}$,
such that $\delta\hat{\Phi}$ and $\delta\hat{\Phi}^{\dagger}$ are
negligible, and we consider $\Phi$ and $\Phi^{*}$ as order parameters.
These parameters describe a macroscopic occupation of the ground state
for particles with spin-$0$ in the presence of self-interaction,
which is represented by the $\lambda$-parameter.

It is worth mentioning that, although our work has a strongly motivation
in possible applications to cosmology and cosmological models, our
proposal is only focused on showing some of the dynamic properties
of RGPE in the presence of an external potential. We emphasize that
a more exhaustive study is required in the field of dark matter and
energy physics in order to elucidate the correspondence with the RGPE.

This paper is organized as follows: In section \ref{sec:Preliminaries}
the variational approach for Klein-Gordon oscillator is briefly introduced.
In section \ref{sec:equation of motion} contains the equations of
motion and their solution, i.e. the stationary solution, the collective
modes, and the free expansion dynamics. The conclusions and outlooks
are sumarized in section \ref{sec:Conclusion}.

\section{Preliminaries: Variational approach}

\label{sec:Preliminaries}

In order to evaluate the collective excitations as well as time-of-flight
dynamics for RGPE, we proposed to extend the variational method with
time-dependent parameters. This method has already proven to be useful
in the studying of trapped Bose-Einstein condensation even in the
presence of vortices \cite{Perez-Garcia-pra56,Teles-pra87,Teles-pra88}.
It consists to write an effective Lagrangian density for a classical
complex field $\Phi\equiv\Phi\left({\bf r},t\right)$, which is given
by 
\begin{equation}
\mathcal{L}=\left(\frac{\partial\Phi^{*}}{\partial t}\right)\left(\frac{\partial\Phi}{\partial t}\right)-\left(\nabla\Phi^{*}\right)\left(\nabla\Phi\right)-\left(m^{2}\omega_{i}^{2}x_{i}^{2}-m\omega_{i}+m^{2}\right)\Phi^{*}\Phi-\frac{\lambda}{Q}\left(\Phi^{*}\Phi\right)^{2}.\label{eq:LD}
\end{equation}
In principle the complex field can be given by 
\begin{equation}
\Phi\left({\bf r},t\right)=\phi\left({\bf r},t\right)e^{i\chi\left({\bf r},t\right)}.
\end{equation}
Let us elucidate some points about this complex field. Different of
the regular Klein-Gordon equation, the wave functions $\Phi\left({\bf r},t\right)$
and $\Phi^{*}\left({\bf r},t\right)$ no longer mean particles with
different charges, because now their product represents the charge
density of a macroscopic state where particles are coherent and indistinguishable.
In other words, since the charge is no longer the difference of particles,
they are impossible to count. Here the density determines the charge
and the phase dynamics determines currents of the trapped relativistic
condensate. Thus, the wave function can be normalized to charge with
the amplitude $\phi\left({\bf r},t\right)$, i.e.
\begin{equation}
Q=\int\phi\left(\mathbf{r},t\right)^{2}d\mathbf{r},
\end{equation}
 and the conserved quantities (currents) are given by
\begin{eqnarray}
j_{0} & = & -\frac{ie}{m}\frac{\int\phi\left(\mathbf{r},t\right)^{2}\frac{\partial\chi\left(\mathbf{r},t\right)}{\partial t}d\mathbf{r}}{\int\phi\left(\mathbf{r},t\right)^{2}d\mathbf{r}},\\
\mathbf{j} & = & -\frac{ie}{m}\frac{\int\phi\left(\mathbf{r},t\right)^{2}\nabla\chi\left(\mathbf{r},t\right)d\mathbf{r}}{\int\phi\left(\mathbf{r},t\right)^{2}d\mathbf{r}}.
\end{eqnarray}
These two quantities are less important if the target is that the
system presents a dynamics. Actually when one works with GPE, the
phase $\chi\left({\bf r},t\right)$ plays a fundamental role which
must be carefully chosen in order to attain physically consistent
results. Nevertheless, in relativistic dynamics we can adopt the gauge
which eliminates the small fluctuations of the phase, and keeps physically
consistent. Therefore, we adopt $\chi=const.$ as the simplest gauge
to work. This is basically an uncharge field.

By following the variational principle, the Lagrangian is calculated
as
\begin{equation}
L\left(t,q_{i},\dot{q}_{i}\right)=\int\mathcal{L}\left(t,x_{i},q_{i},\dot{q}_{i}\right)dx_{i},\label{eq:L}
\end{equation}
which yields the Euler-Lagrange equations 
\begin{equation}
\frac{\partial}{\partial t}\left(\frac{\partial L}{\partial\dot{q}}\right)-\frac{\partial L}{\partial q}=0,
\end{equation}
where $q_{i}\equiv q_{i}\left(t\right)$ are the parameters of the
amplitude $\phi$, and dot is related with the time derivative.

The variational method needs a suitable Ansatz which describes the
ground state of RGPE, such that should be similar to the ground state
of the relativistic Klein-Gordon oscillator (KGO) in limit of low
interaction strength, i.e. $\lambda\ll1$.

\section{Equation of motion for a spherical ground state}

\label{sec:equation of motion}

Since the ground state of KGO can be expressed by a Gaussian function,
we use this one normalized to the charge $Q$ with the center-of-mass
displaced by $\eta_{i}\left(t\right)$ from the center of the harmonic
potential, and the Gaussian width is $\sigma_{i}\left(t\right)=\sigma\left(t\right)$.
This function is given by
\begin{equation}
\phi\left(t,x_{i}\right)=\sqrt{\frac{Q}{\pi^{3/2}\sigma^{3}}}\prod_{i=1}^{3}e^{-\left(x_{i}-\eta_{i}\right)^{2}/2\sigma^{2}}.\label{eq:trialF}
\end{equation}
For those who are acquainted with GPE, this Ansatz is so known as
ideal gas. We choose it instead Thomas-Fermi Ansatz due to the system
has a weak interaction strength when compared with kinetic energy.

By substituting \eqref{eq:trialF} in \eqref{eq:L} and changing to
a dimensionless scale ($\sigma\left(t\right)\rightarrow a_{0}\sigma\left(t\right)$,
$\eta_{i}\left(t\right)\rightarrow a_{0}\eta_{i}\left(t\right)$,
and $\tau\rightarrow\omega t$), we obtain the Lagrangian for a spherical
field
\begin{equation}
L=\frac{Q}{a_{0}^{2}}\left(3-\frac{1}{\alpha^{2}}\right)+\frac{3Q}{2a_{0}^{2}}\left\{ \alpha^{2}\frac{\dot{\sigma}^{2}}{\sigma^{2}}-\frac{1}{\sigma^{2}}-\sigma^{2}-\frac{\gamma}{\sigma^{3}}+\frac{1}{3}\sum_{i=1}^{3}\left(\alpha^{2}\frac{\dot{\eta}_{i}^{2}}{\sigma^{2}}-2\eta_{i}^{2}\right)\right\} ,\label{eq:L2}
\end{equation}
where the oscillator length is $a_{0}=1/\sqrt{m\omega}$, the dimensionless
interaction strength $\gamma=\lambda/3\left(2\pi\right)^{3/2}a_{0}$,
and $\alpha=\sqrt{\omega/m}$ is a dimensionless constant. The Euler-Lagrange
equations results in three equations for the center of mass 
\begin{equation}
\alpha^{2}\left(\ddot{\eta}_{i}-\frac{2\dot{\eta}_{i}\dot{\sigma}}{\sigma}\right)+2\eta_{i}\sigma^{2}=0,\label{eq:com}
\end{equation}
and a fourth equation for the Gaussian width 
\begin{equation}
\alpha^{2}\left(\ddot{\sigma}-\frac{\dot{\sigma}^{2}}{\sigma}+\frac{\sum_{i=1}^{3}\dot{\eta}_{i}^{2}}{3\sigma}\right)+\sigma^{3}=\frac{1}{\sigma}+\frac{3\gamma}{2\sigma^{2}}.\label{eq:sigma}
\end{equation}
Then we can keep the term proportional to $\gamma$ since the ground-state
will not change much, besides just of a scale factor that can be adjusted.
The both above equations of motion can be interpretated as Newton's
equations where each term has a force interpretation such as: quantum
pressure ($1/\sigma$), harmonic confinement ($\sigma^{3}$ and $2\eta_{i}\sigma^{2}$),
and self-interaction ($3\gamma/2\sigma^{2}$). These terms are slightly
different in the context of GPE, however they have the same meaning.
The square velocities and crossed velocities terms are related with
either the buoyancy or the drag force. This viscosity behavior can
be a part of the quantum pressure due to the physical vacuum. By doing
the accelerations and velocities equal to zero ($\ddot{\sigma}=\ddot{\eta}_{i}=\dot{\sigma}=\dot{\eta}_{i}=0$),
we obtain the equilibrium points being a trivial value for the center
of mass 
\begin{equation}
\eta_{0i}=0,
\end{equation}
and a polynomial equation of fifth order for the width 
\begin{equation}
\sigma_{0}^{5}-\sigma_{0}=\frac{3\gamma}{2\sigma_{0}}.\label{eq:ss}
\end{equation}
The solution of eq. \eqref{eq:ss} is found numerically by using the
Newton's method. This is exactly the same stationary solution for
the width equation that comes from GPE.

\subsection{Dipolar and monopolar vibrational modes}

The behavior of low-lying collective oscillations are a consequence
of small perturbations around the stationary condition in trapped
condensates, thus they represent the linear oscillatory modes of the
system. Their collective behavior is due to the interaction strength
described by the non-linear part from RGPE.

By introducing the deviations from equilibrium points such as 
\begin{eqnarray}
\eta_{i}\left(\tau\right) & \approx & \eta_{0i}+\delta\eta_{i}\left(\tau\right),\\
\sigma\left(\tau\right) & \approx & \sigma_{0}+\delta\sigma\left(\tau\right),
\end{eqnarray}
these ones are expanded in Taylor's series until first order terms.
Thus we obtain two uncoupled equations, where the frequencies of the
dipole mode is given by 
\begin{eqnarray}
\alpha^{2}\varpi_{d}^{2} & = & 2\sigma_{0}^{2},
\end{eqnarray}
and the frequency of monopole mode is 
\begin{equation}
\alpha^{2}\varpi_{m}^{2}=4\sigma_{0}^{2}-\frac{3\gamma}{2\sigma_{0}^{3}}.
\end{equation}
Note that the dimensionless frequencies of modes are given by $\varpi_{i}=\omega_{i}/\omega$.

It is worth mentioning that this monopole mode is degenerate, i.e.
there are at least more two quadrupole modes with almost the same
frequency. These quadrupole modes can be calculated by using a non-symmetrical
trial function. This degeneracy would be a subject to explore as next
work. 

\subsection{Nonlinear coupling between dipolar and monopolar modes}

\begin{figure}
\subfloat[Solid black line corresponds to the oscillation of the radius, and
dashed blue line corresponds to oscillation of the center of mass.]{\includegraphics[height=9cm]{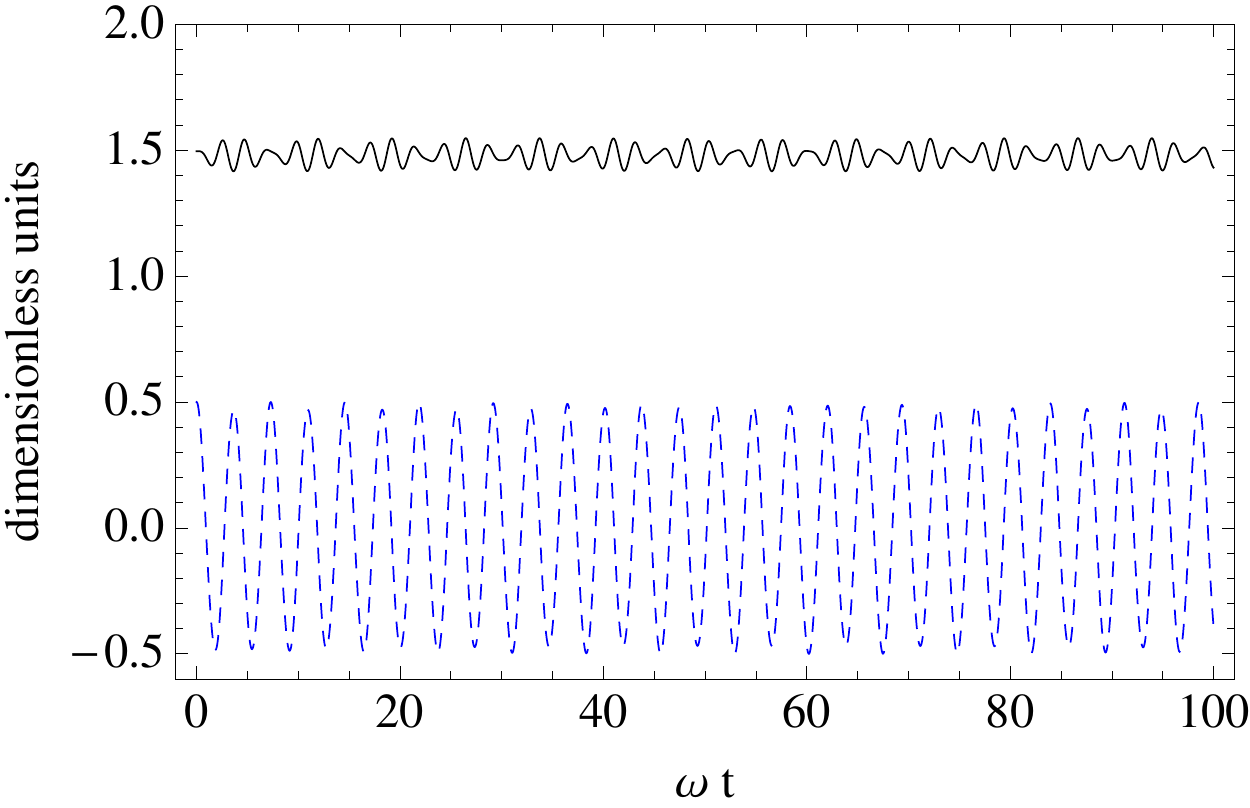}

}

\subfloat[Spectrum of frequencies for the oscillation of the radius.]{\includegraphics[height=5.5cm]{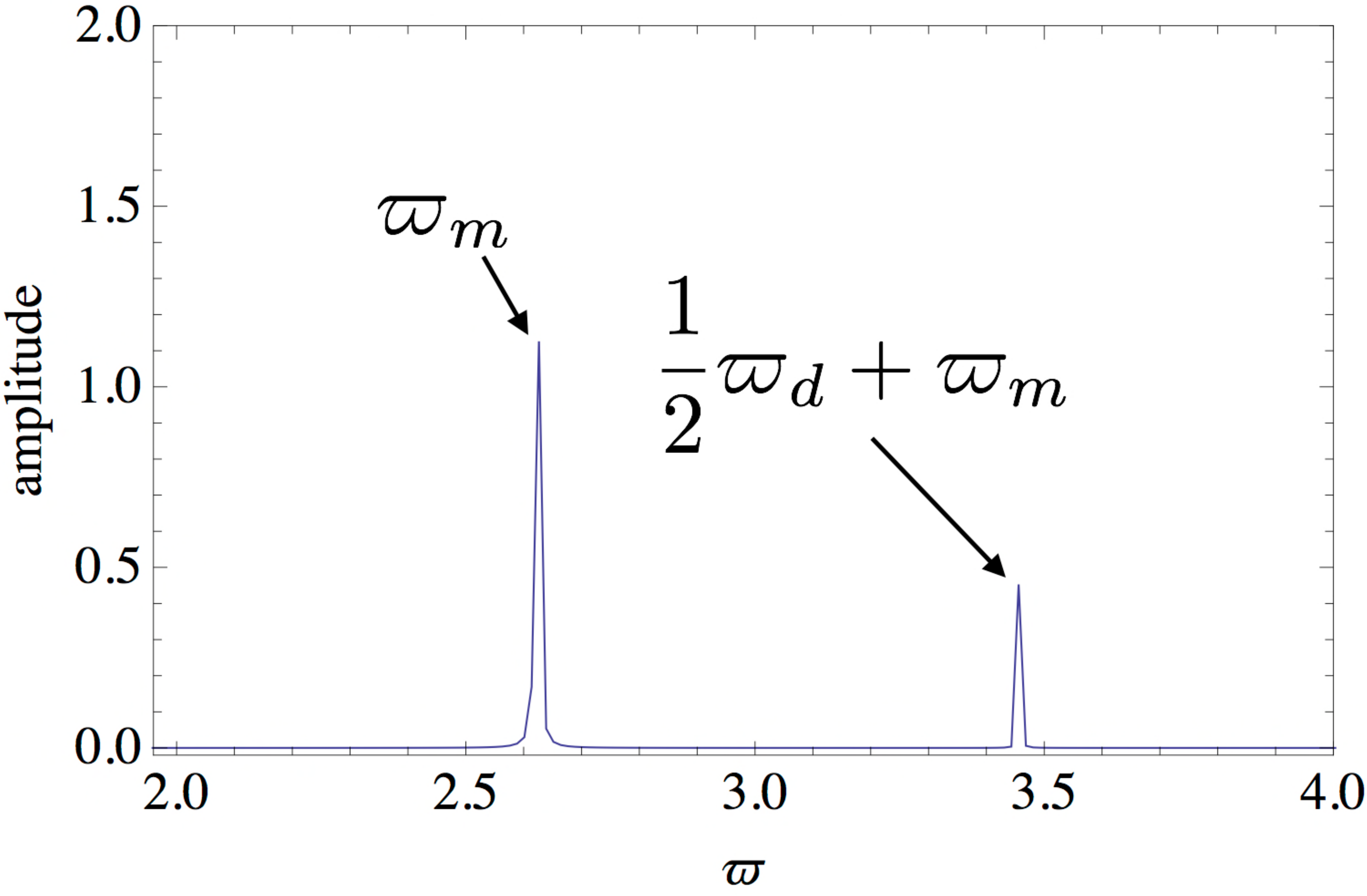}

}\subfloat[Spectrum of frequencies for the oscillation of the center of mass.]{\includegraphics[height=5.5cm]{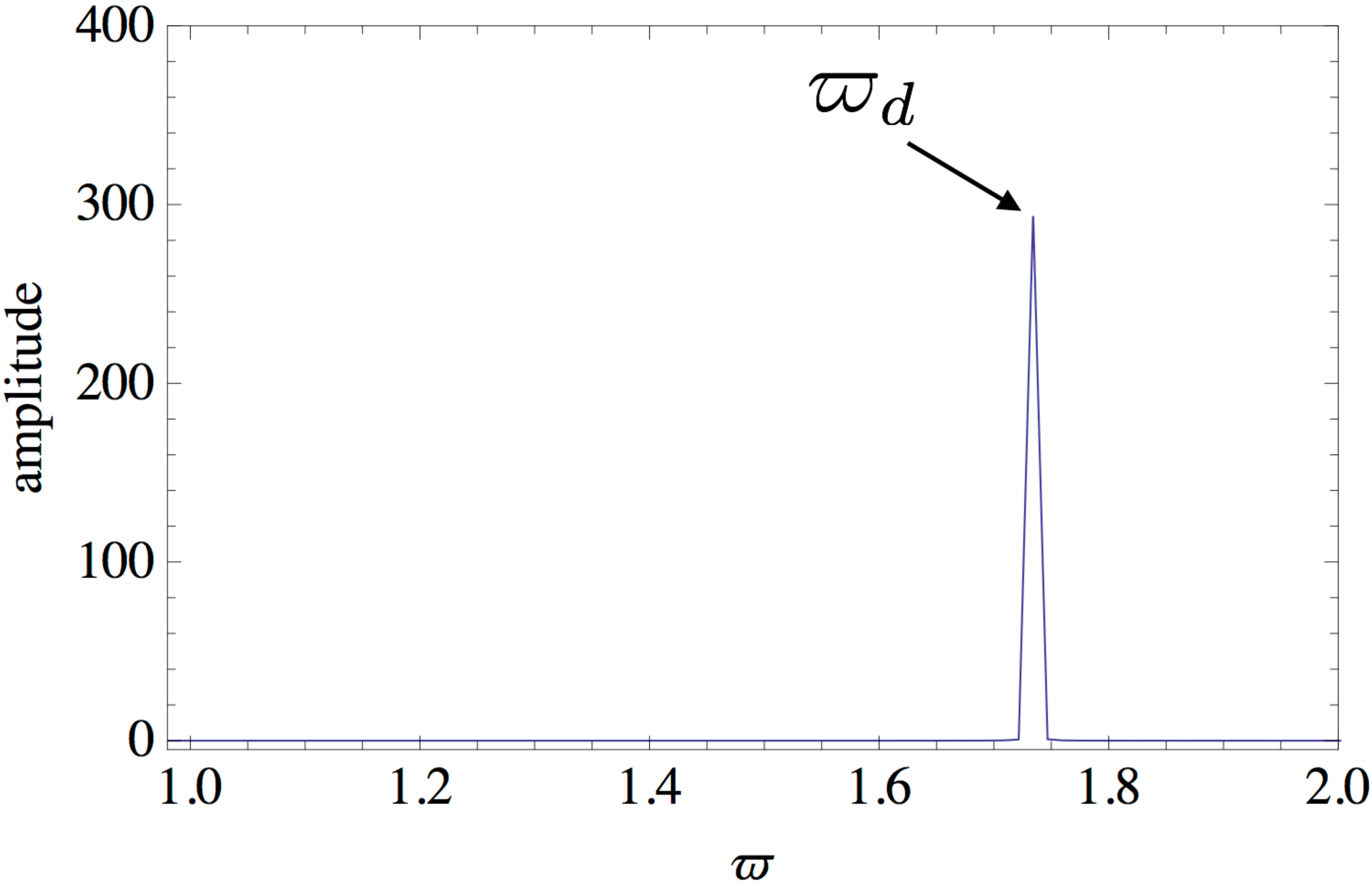}

}

\caption{(Color online) We used $\alpha=1$, $\gamma=1$, $\eta_{01}=\eta_{02}=0$
and $\eta_{03}=0.5$. }

\label{fig:beat}
\end{figure}

Now that we know how the condensate must respond to small perturbations,
we go to solve numerically the nonlinear equations of motion \eqref{eq:com}
and \eqref{eq:sigma}, which can be solved by using the fourth order
Runge-Kutta method. Our initial condition is given by $\eta_{03}=0$,
and $\sigma=\sigma_{0}+\delta\sigma$; where $\delta\sigma$ is a
deviation from equilibrium configuration. Thus we observe that $\sigma$
oscillates with $\omega_{m}$, and the center of mass does not oscillate.
Nevertheless, if we use $\eta_{03}\neq0$ and $\sigma_{0}$ as initial
condition, then we observe a beat in both oscillatory motions as shown
in fig.\ref{fig:beat}. Therefore the motion of the center of mass
is coupled with monopolar mode, where the monopolar mode can be excited
due to the motion of the center of mass, however the opposite situation
is not true. It is a remarkable result because such motions are not
coupled at all for GPE \cite{Perez-Garcia-pra56} -- the motion of
the center of mass (dipolar mode) depends only on harmonic potential
as a consequence of the generalized Ehrenfest theorem, while other
ones such as monopolar and quadrupolar modes are consequence of the
nonlinear effects due to the macroscopic occupation.

The difference between classical and relativistic cases is basically
a second time derivative in the equation \eqref{eq:02}. This time
derivative is responsible for coupling of both modes. Maybe, we can
say that $\alpha^{2}$ is the coupling constant, since this dimensionless
parameter delimits how relativistic the system is. This is also responsible
for the center of mass oscillates with a distinct frequency than the
oscillator.

\subsection{The free expansion dynamics}

\begin{figure}
\includegraphics[height=9cm]{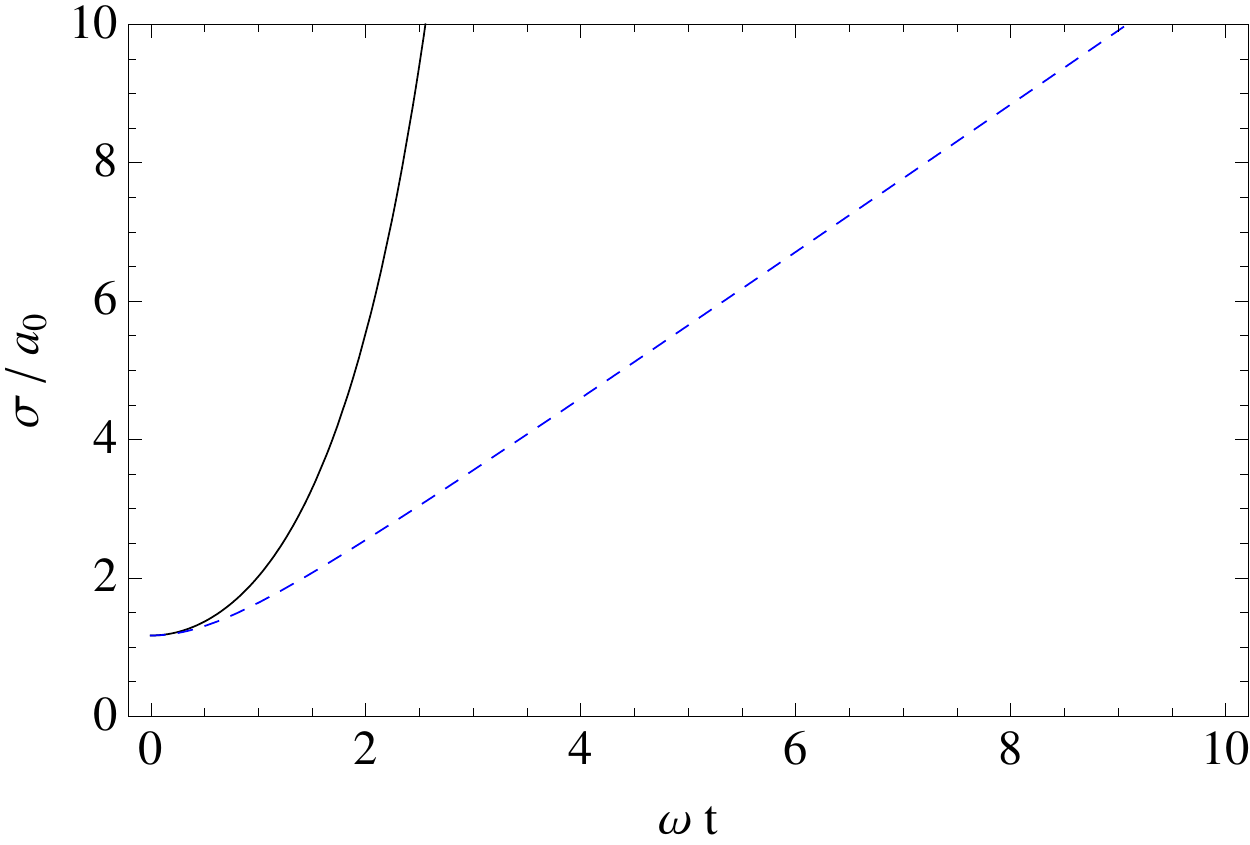}

\caption{(Color online) Comparison of the time-of-flight dynamics between RGPE
(solid black line) and GPE (blue dashed line). We used $\alpha=1$,
$\lambda/\left[2\left(2\pi\right)^{3/2}a_{0}\right]=\left(\sqrt{2/\pi}Na_{s}/a_{0}\right)=1$,
and $\eta_{0i}=0$.}

\label{fig:comparison}
\end{figure}

The time-of-flight pictures constitute the most of common method to
measure an usual BEC. This method consists in switching off the harmonic
trap and letting the atomic cloud expand freely for some time, typically
in order of tens of milliseconds, and then taking pictures of the
expanded cloud \cite{Szczepkowski-revsciinst80}. By cutting off the
terms responsible for the trap in equations \eqref{eq:com} and \eqref{eq:sigma},
we are basically switching off the harmonic confinement. Thus the
equations of motion become: 
\begin{equation}
\ddot{\eta}_{i}-\frac{2\dot{\eta}_{i}\dot{\sigma}}{\sigma}=0,\label{eq:TOFcom}
\end{equation}
for the center of mass, and 
\begin{equation}
\alpha^{2}\left(\ddot{\sigma}-\frac{\dot{\sigma}^{2}}{\sigma}+\frac{\sum_{i=1}^{3}\dot{\eta}_{i}^{2}}{3\sigma}\right)=\frac{1}{\sigma}+\frac{3\gamma}{2\sigma^{2}},\label{eq:TOFsigma}
\end{equation}
for the motion of the condensate width. These equations are numerically
solved by fourth-order Runge-Kutta method. By using the equilibrium
situation as initial condition for the free expansion, we obtain that
RGPE results in an hyper-ballistic free expansion when compared with
the free expansion from GPE (fig. \ref{fig:comparison}). The RGPE
condensate expands too fast that the width reaches a gigantic size
in few milliseconds of free expansion, while GPE results in a ballistic
free expansion. This hyper-ballistic expansion is a consequence of
the viscosity behavior discussed before. The term $\dot{\sigma}^{2}/\sigma$
works as a buoyancy force that impulses the expansion even further.
Nevertheless, its behavior is completely different if the condensate
is oscillating before that the confinement is switched off.

Let us consider that the center of mass is oscillating, consequently
the width is also oscillating, and the confinement is switched off
at time $\tau_{0}=\omega t_{0}$. Thus we find the time evolution
of the velocity of center of mass by integrating equation \eqref{eq:TOFcom}
in time coordinate, which is given by: 
\begin{equation}
\dot{\eta}_{i}\left(\tau\right)=\frac{\dot{\eta}_{i}\left(\tau_{0}\right)}{\sigma\left(\tau_{0}\right)^{2}}\sigma\left(\tau\right)^{2}.
\end{equation}
The above velocity can be replaced in equation \eqref{eq:TOFsigma},
which yields an artificial harmonic confinement as we can note: 
\begin{equation}
\alpha^{2}\left(\ddot{\sigma}-\frac{\dot{\sigma}^{2}}{\sigma}+\frac{\sum_{i=1}^{3}\dot{\eta}_{i}\left(\tau_{0}\right)^{2}}{3\sigma\left(\tau_{0}\right)^{4}}\sigma^{3}\right)=\frac{1}{\sigma}+\frac{3\gamma}{2\sigma^{2}}.
\end{equation}
This artificial confinement has its own equilibrium point which is
calculated by
\begin{equation}
\alpha^{2}\left[\frac{\sum_{i=1}^{3}\dot{\eta}_{i}\left(\tau_{0}\right)^{2}}{3\sigma\left(\tau_{0}\right)^{4}}\right]\sigma_{tof}^{5}-\sigma_{tof}=\frac{3\gamma}{2},
\end{equation}
where the frequency of its linear mode is calculated by introducing
a small deviation in condensate width 
\begin{equation}
\sigma\left(\tau\right)=\sigma_{tof}+\delta\sigma\left(\tau\right).
\end{equation}
Therefore the time-of-flight frequency is given by: 
\begin{eqnarray}
\alpha^{2}\varpi_{tof}^{2} & = & 4\alpha^{2}\left[\frac{\sum_{i=1}^{3}\dot{\eta}_{i}\left(\tau_{0}\right)^{2}}{3\sigma\left(\tau_{0}\right)^{4}}\right]\sigma_{tof}^{2}+\frac{3\gamma}{2\sigma_{tof}^{3}}.
\end{eqnarray}
The artificial confinement is a directly consequence of the second
time derivative on RGPE. Thus we can consider that this is a relativistic
effect of the vacuum viscosity, and we name it as relativistic confinement.
It can also be interpreted as a competition between the buoyancy force
and the drag force.\textbf{ }

\begin{figure}
\centering

\subfloat[Motion of the condensate radius.]{\includegraphics[height=9cm]{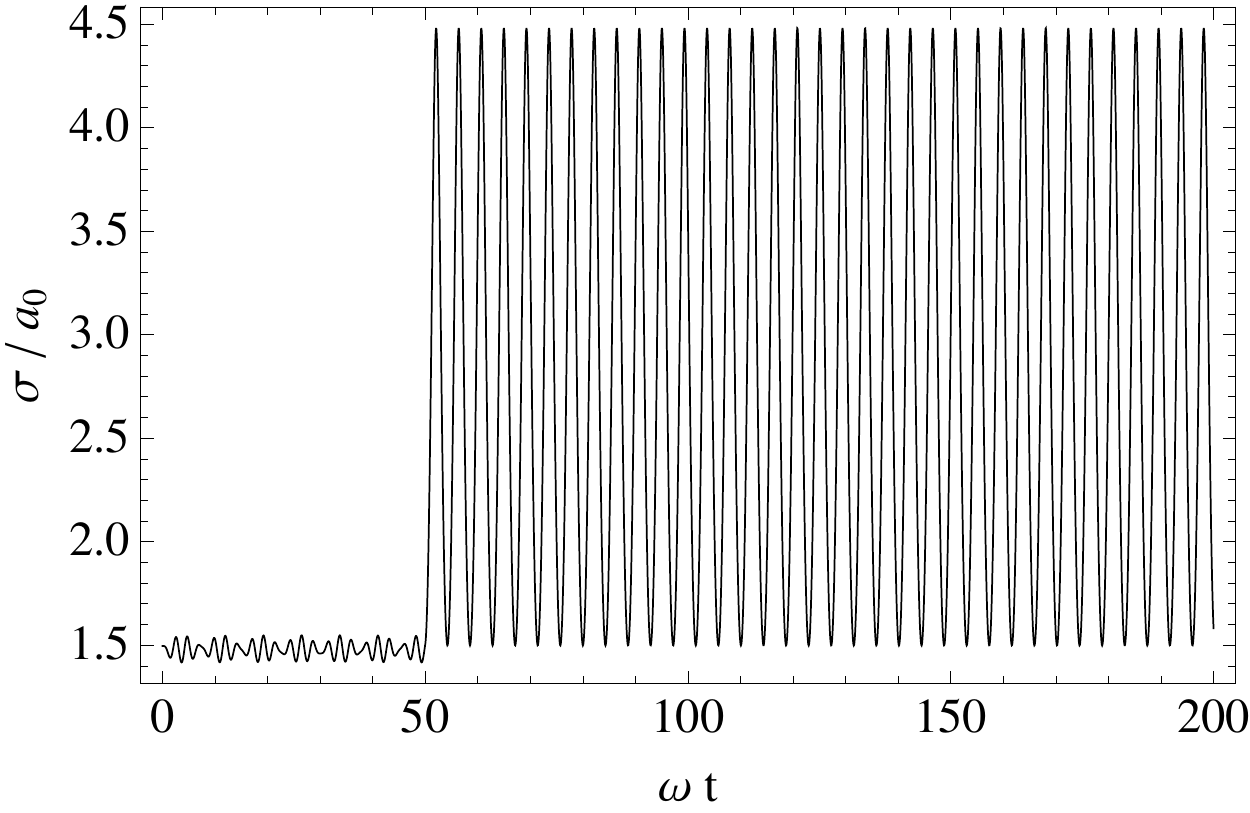}}

\subfloat[Spectrum of frequencies for the motion of the condensate radius.]{\includegraphics[height=5.5cm]{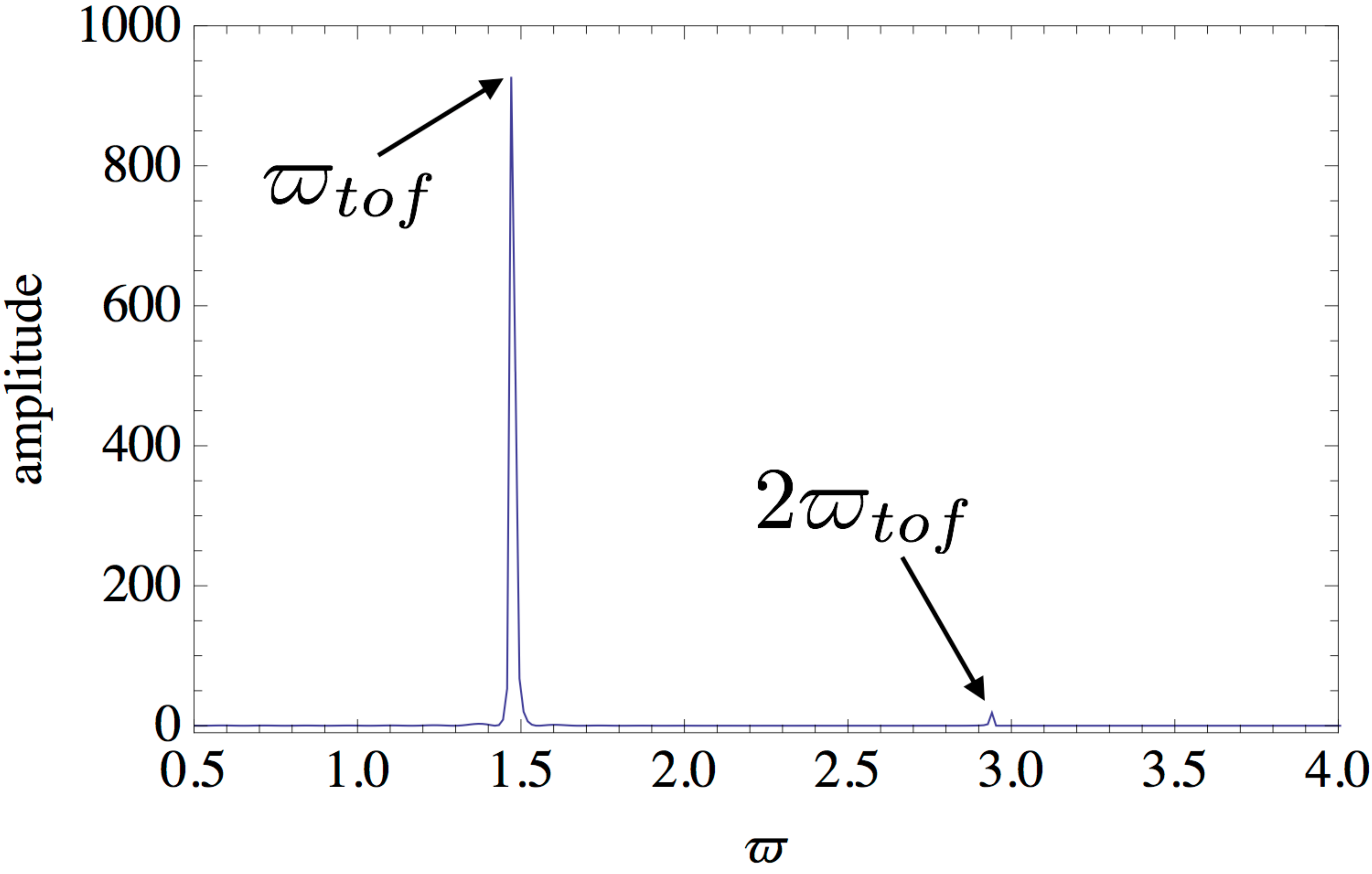}

}\subfloat[Motion of the center of mass.]{\includegraphics[height=5.5cm]{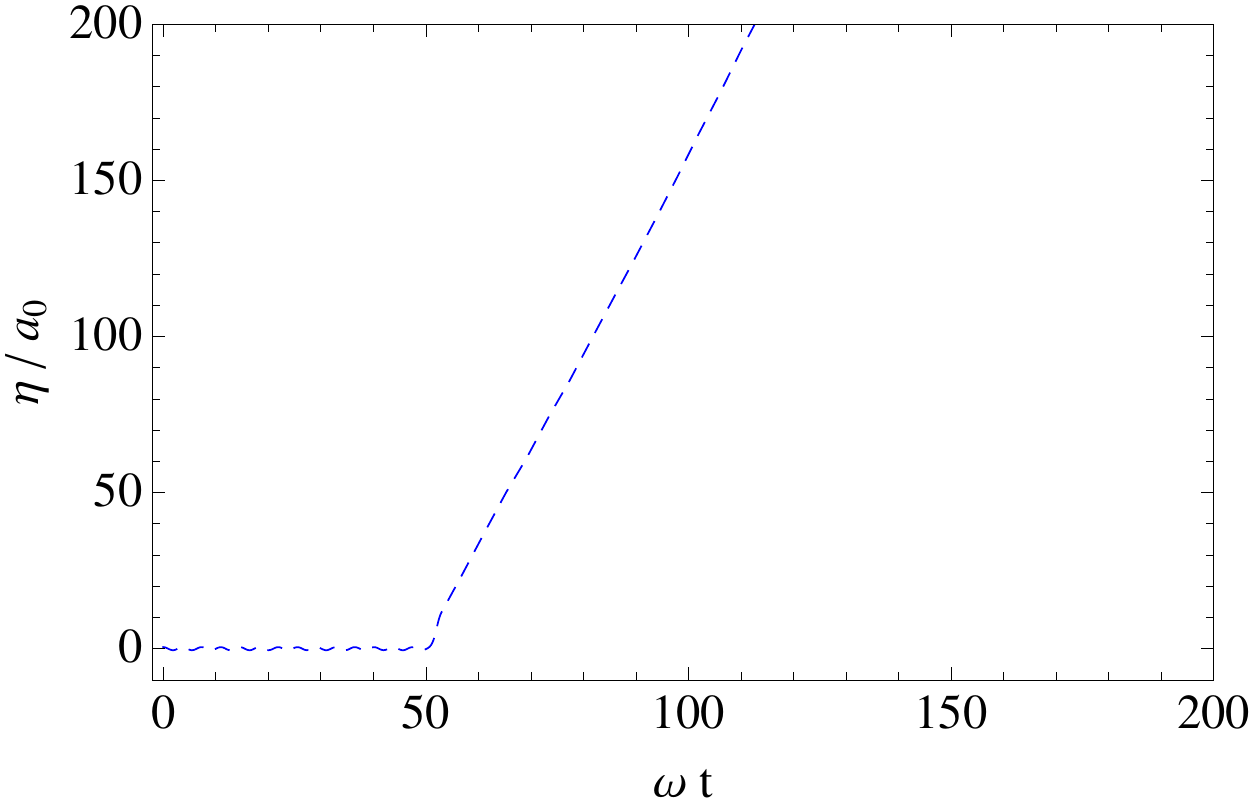}

}

\caption{(Color online) We used $\omega t_{0}=50$, $\alpha=1$, $\gamma=1$,
$\eta_{01}=\eta_{02}=0$ and $\eta_{03}=0.5$.}
\end{figure}

\subsection{Non-relativistic and ultra-relativistic limits}

The regimes present in this system is related with the constant 
\begin{equation}
\alpha^{2}=\frac{\hbar\omega}{mc^{2}}.
\end{equation}
This is a constant that has a ratio of two energy types. The energy
as divisor is clearly the vacuum energy given by the rest mass. In
the numerator one has $\hbar\omega$ which may be related either with
photon energy, or with harmonic potential energy. If $\hbar\omega$
comes from harmonic potential, then $\alpha^{2}$ is interpreted as
a constant that determines the degree of confinement. Indeed, it makes
sense to explain the limits of $\alpha^{2}$ ($\alpha\rightarrow0$,
and $\alpha\rightarrow\infty$), however the equations give us a different
evidence. Since $\alpha^{2}$ appears only in the time-derivatives
and the rest mass energy, it means that $\hbar\omega$ is related
with momentum (i.e. photon energy). Thus $\alpha^{2}$ says how relativistic
is our system.

The non-relativistic case happens when $\alpha\rightarrow0$, because
the equations tends to stationary solution as well as the frequencies
of collective modes tends to infinity. In other words, the classical
stationary case or the higher confinement situation (if $\alpha^{2}$
is interpreted as degree of confinement).

In another hand, the ultra-relativistic case happens when $\alpha\rightarrow\infty$,
because the equations becomes dependent on only accelerations and
velocities:
\begin{eqnarray}
\ddot{\eta}_{i}-\frac{2\dot{\eta}_{i}\dot{\sigma}}{\sigma} & = & 0,\\
\ddot{\sigma}-\frac{\dot{\sigma}^{2}}{\sigma}+\frac{\sum_{i=1}^{3}\dot{\eta}_{i}^{2}}{3\sigma} & = & 0.
\end{eqnarray}
Thus the relativistic confinement is a fraction of ultra-relativistic
confinement in addiction of the self-interaction effect, i.e. $\omega_{tof}^{2}=\frac{4}{3}\omega_{ur}^{2}+\frac{3\gamma}{2\alpha^{2}\sigma_{tof}^{3}}$,
where
\begin{equation}
\lim_{\alpha\rightarrow\infty}\omega_{tof}^{2}=\frac{4}{3}\omega_{ur}^{2}=\frac{4}{3}\left[\frac{\sum_{i=1}^{3}\dot{\eta}_{i}\left(\tau_{0}\right)^{2}}{\sigma\left(\tau_{0}\right)^{4}}\right]\sigma_{tof}^{2}.
\end{equation}
This is the lower confinement situation in the another interpretation.
The system is relativistic for any other value of $\alpha$.

\section{Conclusions}

\label{sec:Conclusion}

In this article, we propose the study of a relativistic version for
the GPE in the presence of a harmonic external potential. The goal
is to point out the differences between RGPE and -- already so well
known -- GPE with respect to the collective excitations as well as
the free expansion dynamics. We assume a mean-field as a macroscopic
occupation of ground state, and try the variational method usually
used with GPE to treat the RGPE.

The spherical harmonic potential is the simplest symmetry to check
out such behaviors. The stationary solution of RGPE is the same of
GPE. The relativistic system shows itself similar in the category
of collective modes. It presents both dipole and monopole modes as
non-relativistic case, however they have a nonlinear coupling unlike
the non-relativistic case. The nonlinearity of the equations allows
that the motion of the center of mass excites the monopole mode, i.e.
the dipole mode transfers kinetic energy to condensate width that
otherwise is not true.

The free expansion is no longer ballistic becoming faster and faster
each time unit. Furthermore, a peculiar expansion behavior appears
when this relativistic entity is evolving a motion of the center of
mass before the confinement shutdown. In this conditions, the condensate
is released with a initial velocity which yields in an artificial
confinement. It means that the condensate oscillates rather than freely
expanding.

Our results suggest the existence of a vacuum viscosity as a consequence
of the relativistic feature. We based our interpretation on the equations
of motion which present forces as either buoyancy-like or drag force-like.
These are opened questions to investigate as next works. It is also
necessary to extend this subject to numerical simulations as well
as a properly study for the dynamic phase. The phase dynamics is able
to reduce the hyper-ballistic behavior, since the system presents
charge-attraction due to the currents.

\begin{acknowledgments}
RPT thanks a financial support from CNPq (Brazil) and FJPC the postdoctoral
fellowship from DGAPA and C\'{a}tedras CONACyT (M\'{e}xico). We
thank to the professor Dr. Kaden Hazzard, its student Ian White, and
the professor Dr. Jorge Seman for their comments.

\end{acknowledgments}

\bibliographystyle{apsrev}
\bibliography{bibkgosc}

\begin{thebibliography}{33}
\expandafter\ifx\csname natexlab\endcsname\relax\def\natexlab#1{#1}\fi
\expandafter\ifx\csname bibnamefont\endcsname\relax
  \def\bibnamefont#1{#1}\fi
\expandafter\ifx\csname bibfnamefont\endcsname\relax
  \def\bibfnamefont#1{#1}\fi
\expandafter\ifx\csname citenamefont\endcsname\relax
  \def\citenamefont#1{#1}\fi
\expandafter\ifx\csname url\endcsname\relax
  \def\url#1{\texttt{#1}}\fi
\expandafter\ifx\csname urlprefix\endcsname\relax\def\urlprefix{URL }\fi
\providecommand{\bibinfo}[2]{#2}
\providecommand{\eprint}[2][]{\url{#2}}

\bibitem[{\citenamefont{Landau}(1941)}]{Landau-physrev60}
\bibinfo{author}{\bibfnamefont{L.}~\bibnamefont{Landau}},
  \bibinfo{journal}{Phys. Rev.} \textbf{\bibinfo{volume}{60}},
  \bibinfo{pages}{356} (\bibinfo{year}{1941}).

\bibitem[{\citenamefont{Abrikosov}(1957)}]{Abrikosov-sovphysJETP5}
\bibinfo{author}{\bibfnamefont{A.~A.} \bibnamefont{Abrikosov}},
  \bibinfo{journal}{Sov. Phys. JETP} \textbf{\bibinfo{volume}{5}},
  \bibinfo{pages}{1174} (\bibinfo{year}{1957}).

\bibitem[{\citenamefont{Nielsen and Olesen}(1973)}]{Nielsen-nuclphysb61}
\bibinfo{author}{\bibfnamefont{H.}~\bibnamefont{Nielsen}} \bibnamefont{and}
  \bibinfo{author}{\bibfnamefont{P.}~\bibnamefont{Olesen}},
  \bibinfo{journal}{Nucl. Phys. B} \textbf{\bibinfo{volume}{61}},
  \bibinfo{pages}{45} (\bibinfo{year}{1973}).

\bibitem[{\citenamefont{Nambu}(1974)}]{Nambu-prd10}
\bibinfo{author}{\bibfnamefont{Y.}~\bibnamefont{Nambu}},
  \bibinfo{journal}{Phys. Rev. D} \textbf{\bibinfo{volume}{10}},
  \bibinfo{pages}{4262} (\bibinfo{year}{1974}).

\bibitem[{\citenamefont{'t~Hooft}(1981)}]{tHooft-nuclphysb190}
\bibinfo{author}{\bibfnamefont{G.}~\bibnamefont{'t~Hooft}},
  \bibinfo{journal}{Nucl. Phys. B} \textbf{\bibinfo{volume}{190}},
  \bibinfo{pages}{455} (\bibinfo{year}{1981}).

\bibitem[{\citenamefont{Rajaraman}(1982)}]{Rajaraman-book1982}
\bibinfo{author}{\bibfnamefont{R.}~\bibnamefont{Rajaraman}},
  \emph{\bibinfo{title}{Solitons and Instantons: An Introduction to Solitons
  and Instantons in Quantum Field Theory}}, North-Holland personal library
  (\bibinfo{publisher}{North-Holland Publishing Company},
  \bibinfo{year}{1982}), ISBN \bibinfo{isbn}{9780444862297}.

\bibitem[{\citenamefont{Peskin and Schroeder}(1995)}]{Peskin-book1995}
\bibinfo{author}{\bibfnamefont{M.}~\bibnamefont{Peskin}} \bibnamefont{and}
  \bibinfo{author}{\bibfnamefont{D.}~\bibnamefont{Schroeder}},
  \emph{\bibinfo{title}{An Introduction to Quantum Field Theory}}, Advanced
  book classics (\bibinfo{publisher}{Addison-Wesley Publishing Company},
  \bibinfo{year}{1995}), ISBN \bibinfo{isbn}{9780201503975}.

\bibitem[{\citenamefont{Gross}(1961)}]{Gross-nc20}
\bibinfo{author}{\bibfnamefont{E.~P.} \bibnamefont{Gross}},
  \bibinfo{journal}{Nuovo Cimento} \textbf{\bibinfo{volume}{20}},
  \bibinfo{pages}{454} (\bibinfo{year}{1961}).

\bibitem[{\citenamefont{Pitaevskii}(1961)}]{Pitaevskii-sovphysJETP13}
\bibinfo{author}{\bibfnamefont{L.~P.} \bibnamefont{Pitaevskii}},
  \bibinfo{journal}{Sov. Phys. JETP} \textbf{\bibinfo{volume}{13}},
  \bibinfo{pages}{451} (\bibinfo{year}{1961}).

\bibitem[{\citenamefont{Gross}(1963)}]{Gross-jmathphys4}
\bibinfo{author}{\bibfnamefont{E.~P.} \bibnamefont{Gross}},
  \bibinfo{journal}{J. Math. Phys.} \textbf{\bibinfo{volume}{4}},
  \bibinfo{pages}{195} (\bibinfo{year}{1963}).

\bibitem[{\citenamefont{Dalfovo et~al.}(1999)\citenamefont{Dalfovo, Giorgini,
  Pitaevskii, and Stringari}}]{Dalfovo-revmodphys71}
\bibinfo{author}{\bibfnamefont{F.}~\bibnamefont{Dalfovo}},
  \bibinfo{author}{\bibfnamefont{S.}~\bibnamefont{Giorgini}},
  \bibinfo{author}{\bibfnamefont{L.~P.} \bibnamefont{Pitaevskii}},
  \bibnamefont{and}
  \bibinfo{author}{\bibfnamefont{S.}~\bibnamefont{Stringari}},
  \bibinfo{journal}{Rev. Mod. Phys.} \textbf{\bibinfo{volume}{71}},
  \bibinfo{pages}{463} (\bibinfo{year}{1999}).

\bibitem[{\citenamefont{Anderson et~al.}(1995)\citenamefont{Anderson, Ensher,
  Matthews, Wieman, and Cornell}}]{Anderson-science269}
\bibinfo{author}{\bibfnamefont{M.~H.} \bibnamefont{Anderson}},
  \bibinfo{author}{\bibfnamefont{J.~R.} \bibnamefont{Ensher}},
  \bibinfo{author}{\bibfnamefont{M.~R.} \bibnamefont{Matthews}},
  \bibinfo{author}{\bibfnamefont{C.~E.} \bibnamefont{Wieman}},
  \bibnamefont{and} \bibinfo{author}{\bibfnamefont{E.~A.}
  \bibnamefont{Cornell}}, \bibinfo{journal}{Science}
  \textbf{\bibinfo{volume}{269}}, \bibinfo{pages}{198} (\bibinfo{year}{1995}).

\bibitem[{\citenamefont{Davis et~al.}(1995)\citenamefont{Davis, Mewes, Andrews,
  van Druten, Durfee, Kurn, and Ketterle}}]{Davis-prl75}
\bibinfo{author}{\bibfnamefont{K.~B.} \bibnamefont{Davis}},
  \bibinfo{author}{\bibfnamefont{M.~O.} \bibnamefont{Mewes}},
  \bibinfo{author}{\bibfnamefont{M.~R.} \bibnamefont{Andrews}},
  \bibinfo{author}{\bibfnamefont{N.~J.} \bibnamefont{van Druten}},
  \bibinfo{author}{\bibfnamefont{D.~S.} \bibnamefont{Durfee}},
  \bibinfo{author}{\bibfnamefont{D.~M.} \bibnamefont{Kurn}}, \bibnamefont{and}
  \bibinfo{author}{\bibfnamefont{W.}~\bibnamefont{Ketterle}},
  \bibinfo{journal}{Phys. Rev. Lett.} \textbf{\bibinfo{volume}{75}},
  \bibinfo{pages}{3969} (\bibinfo{year}{1995}).

\bibitem[{\citenamefont{Bradley et~al.}(1997)\citenamefont{Bradley, Sacket, and
  Hulet}}]{Bradley-prl78}
\bibinfo{author}{\bibfnamefont{C.~C.} \bibnamefont{Bradley}},
  \bibinfo{author}{\bibfnamefont{C.~A.} \bibnamefont{Sacket}},
  \bibnamefont{and} \bibinfo{author}{\bibfnamefont{R.~G.} \bibnamefont{Hulet}},
  \bibinfo{journal}{Phys. Rev. Lett.} \textbf{\bibinfo{volume}{78}},
  \bibinfo{pages}{985} (\bibinfo{year}{1997}).

\bibitem[{\citenamefont{Bloch et~al.}(2008)\citenamefont{Bloch, Dalibard, and
  Zwerger}}]{Bloch-revmodphys80}
\bibinfo{author}{\bibfnamefont{I.}~\bibnamefont{Bloch}},
  \bibinfo{author}{\bibfnamefont{J.}~\bibnamefont{Dalibard}}, \bibnamefont{and}
  \bibinfo{author}{\bibfnamefont{W.}~\bibnamefont{Zwerger}},
  \bibinfo{journal}{Rev. Mod. Phys.} \textbf{\bibinfo{volume}{80}},
  \bibinfo{pages}{885} (\bibinfo{year}{2008}).

\bibitem[{\citenamefont{Fukuyama and Morikawa}(2006)}]{Fukuyama-progthphys115}
\bibinfo{author}{\bibfnamefont{T.}~\bibnamefont{Fukuyama}} \bibnamefont{and}
  \bibinfo{author}{\bibfnamefont{M.}~\bibnamefont{Morikawa}},
  \bibinfo{journal}{Prog. Theor. Phys.} \textbf{\bibinfo{volume}{115}},
  \bibinfo{pages}{1047} (\bibinfo{year}{2006}).

\bibitem[{\citenamefont{Das and Bhaduri}(2015)}]{Das-classquangrav32}
\bibinfo{author}{\bibfnamefont{S.}~\bibnamefont{Das}} \bibnamefont{and}
  \bibinfo{author}{\bibfnamefont{R.~K.} \bibnamefont{Bhaduri}},
  \bibinfo{journal}{Classical and Quantum Gravity}
  \textbf{\bibinfo{volume}{32}}, \bibinfo{pages}{105003}
  (\bibinfo{year}{2015}).

\bibitem[{\citenamefont{Su\'{a}rez and Chavanis}(2015)}]{Suarez-prd92}
\bibinfo{author}{\bibfnamefont{A.}~\bibnamefont{Su\'{a}rez}} \bibnamefont{and}
  \bibinfo{author}{\bibfnamefont{P.-H.} \bibnamefont{Chavanis}},
  \bibinfo{journal}{Phys. Rev. D} \textbf{\bibinfo{volume}{92}},
  \bibinfo{pages}{023510} (\bibinfo{year}{2015}).

\bibitem[{\citenamefont{Pi\~{n}eiro Orioli
  et~al.}(2015)\citenamefont{Pi\~{n}eiro Orioli, Boguslavski, and
  Berges}}]{Pineiro-prd92}
\bibinfo{author}{\bibfnamefont{A.}~\bibnamefont{Pi\~{n}eiro Orioli}},
  \bibinfo{author}{\bibfnamefont{K.}~\bibnamefont{Boguslavski}},
  \bibnamefont{and} \bibinfo{author}{\bibfnamefont{J.}~\bibnamefont{Berges}},
  \bibinfo{journal}{Phys. Rev. D} \textbf{\bibinfo{volume}{92}},
  \bibinfo{pages}{025041} (\bibinfo{year}{2015}).

\bibitem[{\citenamefont{Hazzard and Mueller}(2011)}]{Hazzard-pra84}
\bibinfo{author}{\bibfnamefont{K.~R.~A.} \bibnamefont{Hazzard}}
  \bibnamefont{and} \bibinfo{author}{\bibfnamefont{E.~J.}
  \bibnamefont{Mueller}}, \bibinfo{journal}{Phys. Rev. A}
  \textbf{\bibinfo{volume}{84}}, \bibinfo{pages}{013604}
  (\bibinfo{year}{2011}).

\bibitem[{\citenamefont{Mirza and Mohadesi}(2004)}]{Mirza-communthphys42}
\bibinfo{author}{\bibfnamefont{B.}~\bibnamefont{Mirza}} \bibnamefont{and}
  \bibinfo{author}{\bibfnamefont{M.}~\bibnamefont{Mohadesi}},
  \bibinfo{journal}{Commun. Theor. Phys.} \textbf{\bibinfo{volume}{42}},
  \bibinfo{pages}{664} (\bibinfo{year}{2004}).

\bibitem[{\citenamefont{Chargui et~al.}(2010)\citenamefont{Chargui, Chetouani,
  and Trabelsi}}]{Chargui-communthphys56}
\bibinfo{author}{\bibfnamefont{Y.}~\bibnamefont{Chargui}},
  \bibinfo{author}{\bibfnamefont{L.}~\bibnamefont{Chetouani}},
  \bibnamefont{and} \bibinfo{author}{\bibfnamefont{A.}~\bibnamefont{Trabelsi}},
  \bibinfo{journal}{Commun. Theor. Phys.} \textbf{\bibinfo{volume}{53}},
  \bibinfo{pages}{231} (\bibinfo{year}{2010}).

\bibitem[{\citenamefont{Moshinsky and
  Szczepaniak}(1989)}]{Moshinsky-jpamaththeo22}
\bibinfo{author}{\bibfnamefont{M.}~\bibnamefont{Moshinsky}} \bibnamefont{and}
  \bibinfo{author}{\bibfnamefont{A.}~\bibnamefont{Szczepaniak}},
  \bibinfo{journal}{J. Phys. A: Math. Gen.} \textbf{\bibinfo{volume}{22}},
  \bibinfo{pages}{L817} (\bibinfo{year}{1989}).

\bibitem[{\citenamefont{Ben\'{i}tez et~al.}(1990)\citenamefont{Ben\'{i}tez,
  Mart\'{i}nez~y Romero, N\'{u}\~{n}ez Y\'{e}pez, and
  Salas-Brito}}]{Benitez-prl64}
\bibinfo{author}{\bibfnamefont{J.}~\bibnamefont{Ben\'{i}tez}},
  \bibinfo{author}{\bibfnamefont{R.~P.} \bibnamefont{Mart\'{i}nez~y Romero}},
  \bibinfo{author}{\bibfnamefont{H.~N.} \bibnamefont{N\'{u}\~{n}ez Y\'{e}pez}},
  \bibnamefont{and} \bibinfo{author}{\bibfnamefont{A.~L.}
  \bibnamefont{Salas-Brito}}, \bibinfo{journal}{Phys. Rev. Lett.}
  \textbf{\bibinfo{volume}{64}}, \bibinfo{pages}{1643} (\bibinfo{year}{1990}).

\bibitem[{\citenamefont{Dom\'{i}nguez-Adame and
  Gonz\'{a}lez}(1990)}]{Dominguez-eurphyslett13}
\bibinfo{author}{\bibfnamefont{F.}~\bibnamefont{Dom\'{i}nguez-Adame}}
  \bibnamefont{and} \bibinfo{author}{\bibfnamefont{M.~A.}
  \bibnamefont{Gonz\'{a}lez}}, \bibinfo{journal}{Europhys. Lett.}
  \textbf{\bibinfo{volume}{13}}, \bibinfo{pages}{193} (\bibinfo{year}{1990}).

\bibitem[{\citenamefont{Dom\'{i}nguez-Adame}(1992)}]{Dominguez-physletta162}
\bibinfo{author}{\bibfnamefont{F.}~\bibnamefont{Dom\'{i}nguez-Adame}},
  \bibinfo{journal}{Phys. Lett. A} \textbf{\bibinfo{volume}{162}},
  \bibinfo{pages}{18} (\bibinfo{year}{1992}).

\bibitem[{\citenamefont{Rao and Kagali}(2008)}]{Rao-physscr77}
\bibinfo{author}{\bibfnamefont{N.~A.} \bibnamefont{Rao}} \bibnamefont{and}
  \bibinfo{author}{\bibfnamefont{B.~A.} \bibnamefont{Kagali}},
  \bibinfo{journal}{Phys. Scr.} \textbf{\bibinfo{volume}{77}},
  \bibinfo{pages}{015003} (\bibinfo{year}{2008}).

\bibitem[{\citenamefont{Boumali et~al.}(2011)\citenamefont{Boumali, Hafdallah,
  and Toumi}}]{Boumali-physscr84}
\bibinfo{author}{\bibfnamefont{A.}~\bibnamefont{Boumali}},
  \bibinfo{author}{\bibfnamefont{A.}~\bibnamefont{Hafdallah}},
  \bibnamefont{and} \bibinfo{author}{\bibfnamefont{A.}~\bibnamefont{Toumi}},
  \bibinfo{journal}{Phys. Scr.} \textbf{\bibinfo{volume}{84}},
  \bibinfo{pages}{037001} (\bibinfo{year}{2011}).

\bibitem[{\citenamefont{Boumali and Chetouani}(2005)}]{Boumali-physletta346}
\bibinfo{author}{\bibfnamefont{A.}~\bibnamefont{Boumali}} \bibnamefont{and}
  \bibinfo{author}{\bibfnamefont{L.}~\bibnamefont{Chetouani}},
  \bibinfo{journal}{Phys. Lett. A} \textbf{\bibinfo{volume}{346}},
  \bibinfo{pages}{261} (\bibinfo{year}{2005}).

\bibitem[{\citenamefont{P\'{e}rez-Garc\'{i}a
  et~al.}(1997)\citenamefont{P\'{e}rez-Garc\'{i}a, Humberto, Cirac, Lewenstein,
  and Zoller}}]{Perez-Garcia-pra56}
\bibinfo{author}{\bibfnamefont{V.~M.} \bibnamefont{P\'{e}rez-Garc\'{i}a}},
  \bibinfo{author}{\bibfnamefont{M.}~\bibnamefont{Humberto}},
  \bibinfo{author}{\bibfnamefont{J.~I.} \bibnamefont{Cirac}},
  \bibinfo{author}{\bibfnamefont{M.}~\bibnamefont{Lewenstein}},
  \bibnamefont{and} \bibinfo{author}{\bibfnamefont{P.}~\bibnamefont{Zoller}},
  \bibinfo{journal}{Phys. Rev. A} \textbf{\bibinfo{volume}{56}},
  \bibinfo{pages}{1424} (\bibinfo{year}{1997}).

\bibitem[{\citenamefont{Teles et~al.}(2013{\natexlab{a}})\citenamefont{Teles,
  dos Santos, Caracanhas, and Bagnato}}]{Teles-pra87}
\bibinfo{author}{\bibfnamefont{R.~P.} \bibnamefont{Teles}},
  \bibinfo{author}{\bibfnamefont{F.~E.~A.} \bibnamefont{dos Santos}},
  \bibinfo{author}{\bibfnamefont{M.~A.} \bibnamefont{Caracanhas}},
  \bibnamefont{and} \bibinfo{author}{\bibfnamefont{V.~S.}
  \bibnamefont{Bagnato}}, \bibinfo{journal}{Phys. Rev. A}
  \textbf{\bibinfo{volume}{87}}, \bibinfo{pages}{033622}
  (\bibinfo{year}{2013}{\natexlab{a}}).

\bibitem[{\citenamefont{Teles et~al.}(2013{\natexlab{b}})\citenamefont{Teles,
  Bagnato, and dos Santos}}]{Teles-pra88}
\bibinfo{author}{\bibfnamefont{R.~P.} \bibnamefont{Teles}},
  \bibinfo{author}{\bibfnamefont{V.~S.} \bibnamefont{Bagnato}},
  \bibnamefont{and} \bibinfo{author}{\bibfnamefont{F.~E.~A.} \bibnamefont{dos
  Santos}}, \bibinfo{journal}{Phys. Rev. A} \textbf{\bibinfo{volume}{88}},
  \bibinfo{pages}{053613} (\bibinfo{year}{2013}{\natexlab{b}}).

\bibitem[{\citenamefont{Szczepkowski et~al.}(2009)\citenamefont{Szczepkowski,
  Gartman, Witkowski, Tracewski, Zawada, and
  Gawlik}}]{Szczepkowski-revsciinst80}
\bibinfo{author}{\bibfnamefont{J.}~\bibnamefont{Szczepkowski}},
  \bibinfo{author}{\bibfnamefont{R.}~\bibnamefont{Gartman}},
  \bibinfo{author}{\bibfnamefont{M.}~\bibnamefont{Witkowski}},
  \bibinfo{author}{\bibfnamefont{L.}~\bibnamefont{Tracewski}},
  \bibinfo{author}{\bibfnamefont{M.}~\bibnamefont{Zawada}}, \bibnamefont{and}
  \bibinfo{author}{\bibfnamefont{W.}~\bibnamefont{Gawlik}},
  \bibinfo{journal}{Rev. Sci. Instrum.} \textbf{\bibinfo{volume}{80}},
  \bibinfo{eid}{053103} (\bibinfo{year}{2009}).

\end{thebibliography}

\end{document}